\documentclass[cits]{PoS}\usepackage{amsmath} % HEPHY-PUB 956/15 (2015)
\title{Strong Couplings of Charmed Mesons and Quarkonia}
\ShortTitle{Strong Couplings of Charmed Mesons and Quarkonia}
\author{Wolfgang Lucha\\Institute for High Energy Physics,
Austrian Academy of Sciences, Nikolsdorfergasse 18, A-1050 Vienna,
Austria\\E-mail: \email{Wolfgang.Lucha@oeaw.ac.at}}
\author{\speaker{Dmitri Melikhov}\\D.~V.~Skobeltsyn Institute of
Nuclear Physics, Moscow State University, 119991, Moscow, Russia,
and\\Faculty of Physics, University of Vienna, Boltzmanngasse 5,
A-1090 Vienna, Austria\\E-mail: \email{dmitri\_melikhov@gmx.de}}
\author{Hagop Sazdjian\\IPN, CNRS/IN2P3, Universit\`e Paris-Sud
11, F-91406 Orsay, France\\E-mail: \email{sazdjian@ipno.in2p3.fr}}
\author{Silvano Simula\\INFN, Sezione di Roma Tre, Via della Vasca
Navale 84, I-00146 Roma, Italy\\E-mail:
\email{simula@roma3.infn.it}}

\abstract{We extract the strong coupling constants of three
mesons, each of which is composed of either two charm quarks or
one charm quark and one light (\emph{i.e.}, $u,$ $d,$ or $s$)
quark, from the matrix elements for the transitions of two of
these mesons induced by appropriate quark currents within the
framework of a relativistic dispersion approach to the
constituent-quark picture of mesons. Among others, we also analyse
the impact of the violation of the SU(3) flavour symmetry by the
quark masses. In the case of mesons containing one light quark, we
observe, in two respects, discrepancies between~our findings and
the predictions of QCD sum rules: our strong couplings exceed
considerably the ones emerging from QCD sum rules, and, in our
approach, the replacement of a light quark by a strange quark
entails, in contrast to QCD sum rules, a reduction of the
magnitudes of the strong couplings.}

\FullConference{The European Physical Society Conference on High
Energy Physics\\22--29 July 2015\\Vienna, Austria}\begin{document}

\section{Lots of definitions: strong couplings, transition form
factors, and decay constants}By use of a relativistic dispersion
approach relying on the constituent-quark model for hadrons, we
analyse the three-meson \emph{strong couplings\/} $g_{PP'V}$ and
$g_{PV'V}$ for vector mesons $V,$ with polarization vectors
$\varepsilon_\mu,$ and pseudoscalar mesons $P$: physical
parameters defined by the three-meson amplitudes
\begin{align*}\langle P'(p_2)\,V(q)|P(p_1)\rangle&=
-\frac{\textcolor{red}{g_{PP'V}}}{2}\,(p_1+p_2)^\mu\,
\varepsilon^*_\mu(q)\ ,\\ \langle V'(p_2)\,V(q)|P(p_1)\rangle
&=-\textcolor{red}{g_{PV'V}}\,
\epsilon_{\varepsilon^*(q)\,\varepsilon^*(p_2)\,p_1\,p_2}\ ,\qquad
q\equiv p_1-p_2\ .\end{align*}Of particular interest to us will be
any system with, at least, one of the charmonia, $J/\psi$ or
$\eta_c,$ among the meson triple. These strong couplings enter
into the residues of poles in corresponding~\emph{transition form
factors\/} for $q^2>0$ arising from intermediate meson states. The
form factors relevant for us~read\begin{align*}\langle
P'(p_2)|\bar q_2\,\gamma_\mu\,q_1|P(p_1)\rangle
&=\textcolor{red}{F_+^{P\to P'}(q^2)}\,(p_1+p_2)_\mu+\cdots\ ,\\
\langle V(p_2)|\bar q_2\,\gamma_\mu\,q_1|P(p_1)\rangle
&=\frac{2\,\textcolor{red}{V^{P\to V}(q^2)}}{M_P+M_V}\,
\epsilon_{\mu\,\varepsilon^*(p_2)\,p_1\,p_2}\ ,\\\langle
V(p_2)|\bar q_2\,\gamma_\mu\,\gamma_5\,q_1|P(p_1)\rangle&={\rm
i}\,q_\mu\left(\varepsilon^*(p_2)\,p_1\right)\frac{2\,M_V}{q^2}
\,\textcolor{red}{A_0^{P\to V}(q^2)}+\cdots\ .\end{align*}The
\emph{decay constants\/} $f_V$ and $f_P$ of vector and
pseudoscalar mesons are defined, by meson-to-vacuum transition
amplitudes of vector quark currents $\bar q_1\,\gamma_\mu\,q_2$
and axial-vector quark currents $\bar
q_1\,\gamma_\mu\,\gamma_5\,q_2,$~by$$\langle0|\bar
q_1\,\gamma_\mu\,q_2|V(q)\rangle=\textcolor{red}{f_V}\,M_V\,
\varepsilon_\mu(q)\ ,\qquad\langle0|\bar
q_1\,\gamma_\mu\,\gamma_5\,q_2|P(q)\rangle={\rm
i}\,\textcolor{red}{f_P}\,q_\mu\ .$$The \emph{poles\/} of the form
factors, residing at pseudoscalar or vector resonances $P_R$ or
$V_R,$ are of the~form
\begin{align*}F_+^{P\to P'}(q^2)&=\frac{g_{PP'V_R}\,f_{V_R}}
{2\,M_{V_R}}\,\frac{1}{1-\frac{q^2}{M_{V_R}^2}}+\cdots\ ,\\
V^{P\to V}(q^2)&=\frac{(M_V+M_P)\,g_{PVV_R}\,f_{V_R}}{2\,M_{V_R}}
\,\frac{1}{1-\frac{q^2}{M^2_{V_R}}}+\cdots\ ,\\[-.7ex]A_0^{P\to
V}(q^2)&=\frac{g_{PP_RV}\,f_{P_R}}{2\,M_{V}}\,
\frac{1}{1-\frac{q^2}{M^2_{P_R}}}+\cdots\ .\end{align*}

\section{Quark-model-based dispersion approach}
\begin{figure}[h]\begin{center}\begin{tabular}{ccccc}
\includegraphics[scale=.2917]{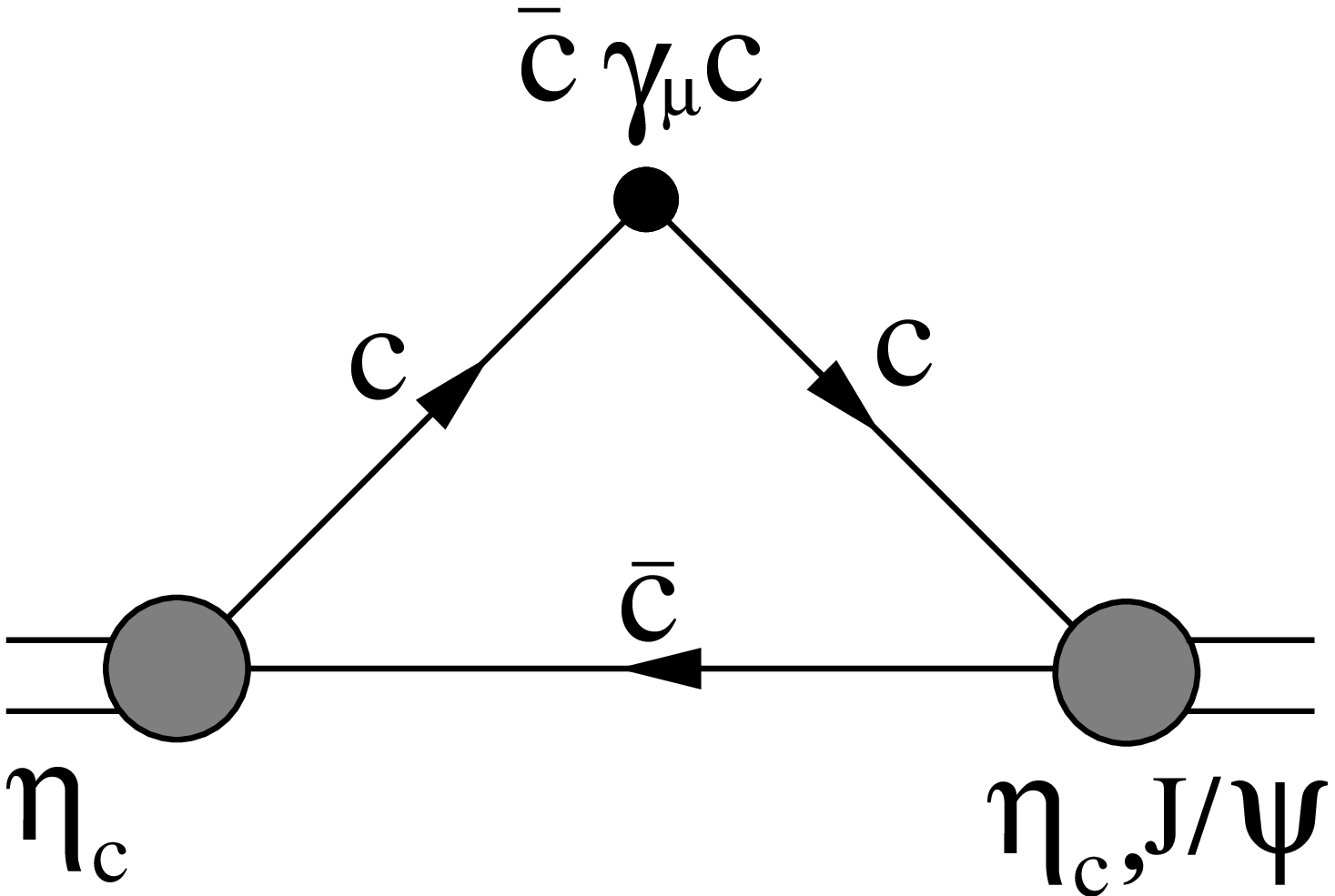}&&
\includegraphics[scale=.2917]{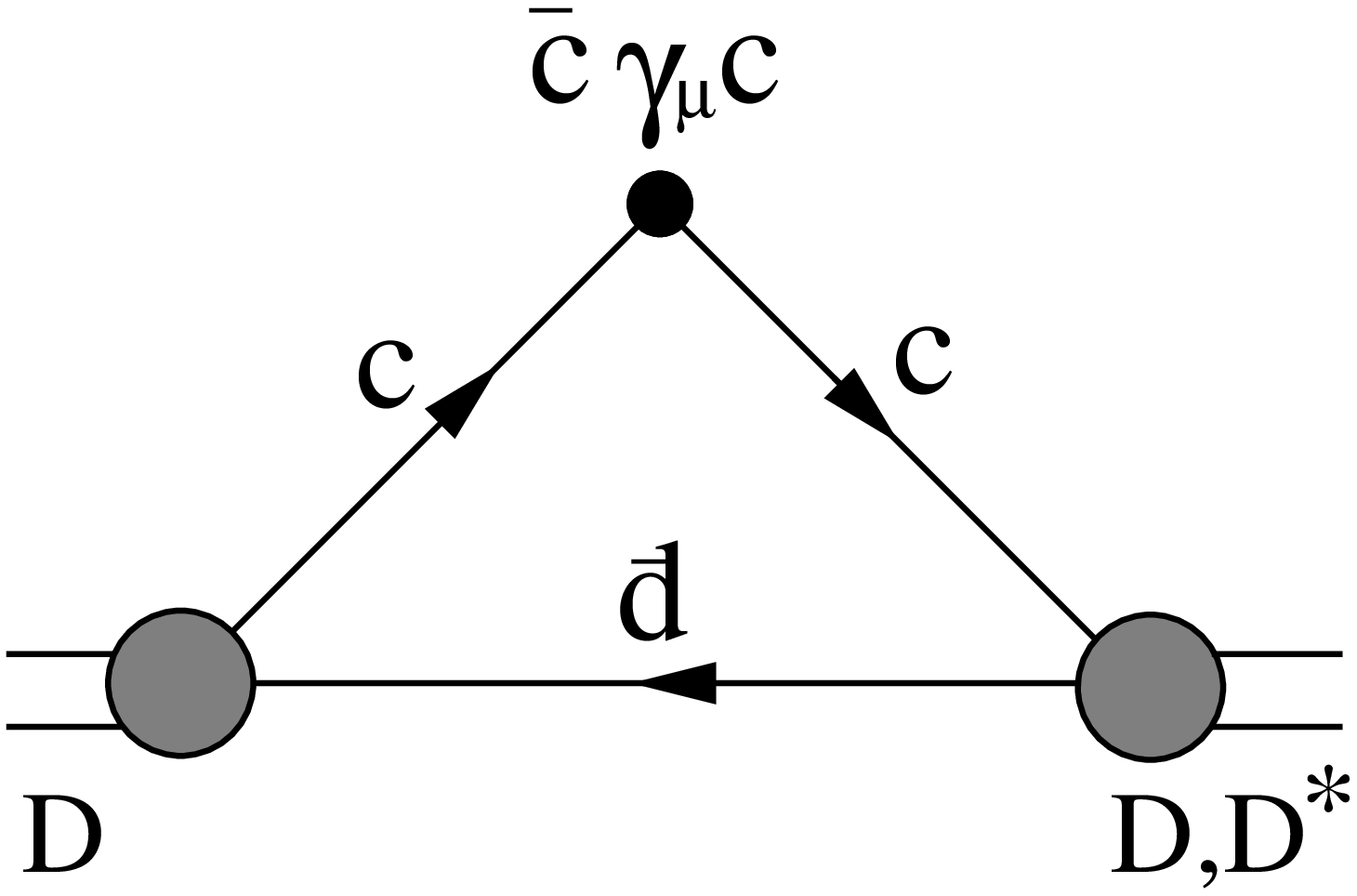}&&
\includegraphics[scale=.2917]{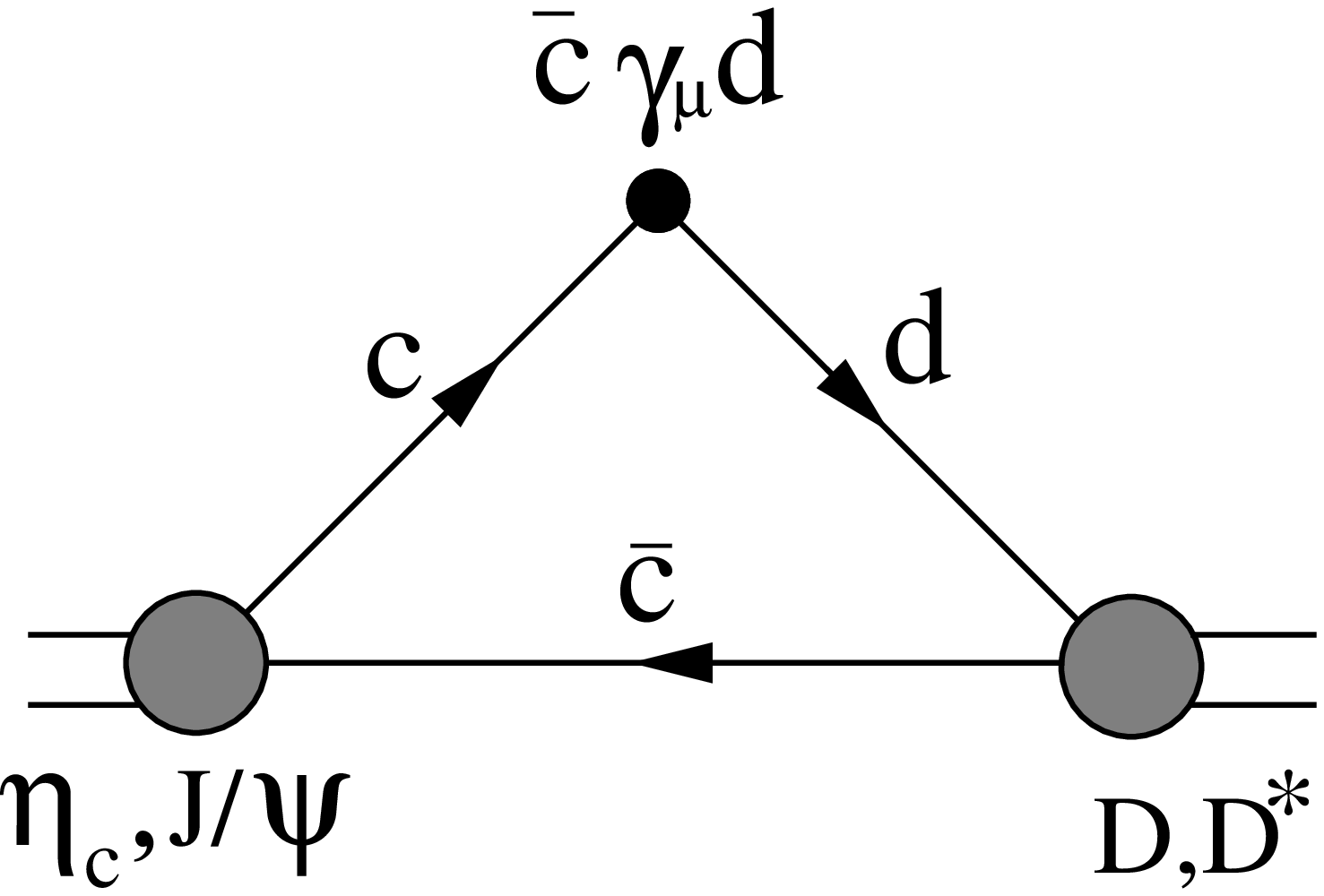}\\(a)&&(b)&&(c)
\end{tabular}\caption{(One-loop) Feynman graphs for transitions
induced by constituent-quark vector currents $\bar
Q_1\,\gamma_\mu\,Q_2$.} \label{Fig:BDC}\end{center}\end{figure}

We calculate the
three relevant form factors, \emph{viz.}, $F_+^{P\to P'}(q^2),$
$V^{P\to V}(q^2),$ and $A_0^{P\to V}(q^2),$ within the framework
of a relativistic \emph{constituent-quark\/} picture \cite{CQM}.
To this end, we must relate the currents defining the form factors
to their constituent-quark ($Q$) counterparts. The task of
establishing such a relationship is easily accomplished for
heavy-quark currents by introducing form factors $g_V$ and~$g_A,$
$$\bar q_1\,\gamma_\mu\,q_2=g_V\,\bar Q_1\,\gamma_\mu\,Q_2+\cdots\
,\qquad\bar q_1\, \gamma_\mu\,\gamma_5\,q_2=g_A\,\bar
Q_1\,\gamma_\mu\,\gamma_5\,Q_2 +\cdots\ ,$$but is not that easy
for the case of light quarks \cite{GVA}. Numerically, we use
\cite{MS} for the constituent-quark masses and couplings
$m_d=m_u=0.23\;\mbox{GeV},$ $m_s=0.35\;\mbox{GeV},$
$m_c=1.45\;\mbox{GeV},$ and $g_V=g_A=1.$ The relativistic
dispersion technique enables us to represent all the quantities of
interest by integrals~over invariant masses of intermediate
quark--antiquark states of \emph{spectral densities\/} derived
from one-loop Feynman diagrams and \emph{wave functions\/} of the
relevant pseudoscalar or vector mesons \cite{RDA} of~the form
\begin{align*}&\phi_{P,V}(s)=\frac{\pi}{\sqrt2}\,
\frac{\sqrt{s^2-(m_1^2-m^2)^2}}{\sqrt{s-(m_1-m)^2}}\,
\frac{w_{P,V}(k^2)}{s^{3/4}}\
,\\&k^2=\frac{(s-m_1^2-m^2)^2-4\,m_1^2\,m^2}{4\,s}\
,\qquad\int{\rm d}k\,k^2\,w^2_{P,V}(k^2)=1\ .\end{align*}For the
radial meson wave functions $w_{P,V}(k^2),$ we'll find sufficient
to assume some Gaussian shapes$$w_{P,V}(k^2)\propto\exp\!\left(
-\frac{k^2}{2\,\beta_{P,V}^2}\right)$$fixed by parameters
$\beta_{P,V}.$ Within the dispersion formalism, the decay
constants $f_{P,V}$ become spectral integrals of densities
$\rho_{P,V}(s),$ and the form factors $F_+^{P\to P'}(q^2),$
$V^{P\to V}(q^2)$ and $A_0^{P\to V}(q^2),$ generically called
${\cal F}(q^2),$ become double dispersion integrals of suitable
double spectral densities $\Delta(s_1,s_2,q^2)$:$$f_{P,V}=\int{\rm
d}s\,\phi_{P,V}(s)\,\rho_{P,V}(s)\ ,\qquad{\cal F}(q^2)=\int{\rm
d}s_1\,\phi_1(s_1)\int{\rm d}s_2\,\phi_2(s_2)\,\Delta(s_1,s_2,q^2)
\ .$$The double spectral densities $\Delta(s_1,s_2,q^2)$ can be
rather straightforwardly derived from the Feynman diagrams shown
in Fig.~\ref{Fig:BDC}. Table~\ref{Tab:MDCS} subsumes the required
parameter values of the mesons discussed.

\begin{table}[h]\begin{center}\caption{Masses $M,$ decay
constants $f_{P,V}$ and slope parameters $\beta_{P,V}$ of the
$D_{(s)}^{(*)}$ mesons and charmonia $c\bar c$~\cite{MP}.}
\label{Tab:MDCS}\vspace{2ex}\begin{tabular}{lcccccc}
\hline\hline\\[-2.7ex]\multicolumn{1}{c}{Meson}&$D$&$D^*$&$D_s$&
$D_s^*$&$\eta_c$&$J/\psi$\\[0.5ex]\hline\\[-2.7ex]$M$ (GeV)\
&$1.87$& $2.010$&$1.97$&$2.11$&$2.980$&$3.097$\\$f_{P,V}$ (MeV)\
&\ $206\pm8$\ &\ $260\pm10$\ &\ $248\pm2.5$\ &\ $311\pm9$\ &\
$394.7\pm2.4$\ &\ $405\pm7$\ \\$\beta_{P,V}$ (GeV)\ &$0.475$&
$0.48$&$0.545$&$0.54$&$0.77$&$0.68$\\[.5ex]\hline\hline
\end{tabular}\end{center}\end{table}

\noindent All form factors ${\cal F},$ computed off the actual
locations of their resonances $R,$ may be \emph{interpolated\/}~by
$${\cal F}(q^2)= \frac{{\cal F}(0)}{\left(1-\frac{q^2}{M_R^2}
\right)\!\left(1-\frac{\sigma_1\,q^2}{M_R^2}+\frac{\sigma_2\,q^4}
{M_R^4}\right)}\ ,\qquad\mbox{Res}\,{\cal F}(M_R^2)=\frac{{\cal
F}(0)}{1-\sigma_1+\sigma_2}\ .$$From the \emph{residues\/} of
these form factors --- involving the meson masses, decay
constants, and strong couplings --- the latter are derived, by
combined fits if they are present in more than one form factor.

\section{Strong couplings of three charmonia:
$\eta_c\,\eta_c\,J/\psi$ and $\eta_c\,J/\psi\,J/\psi$}
\begin{figure}[h]\begin{center}
\includegraphics[scale=.70742]{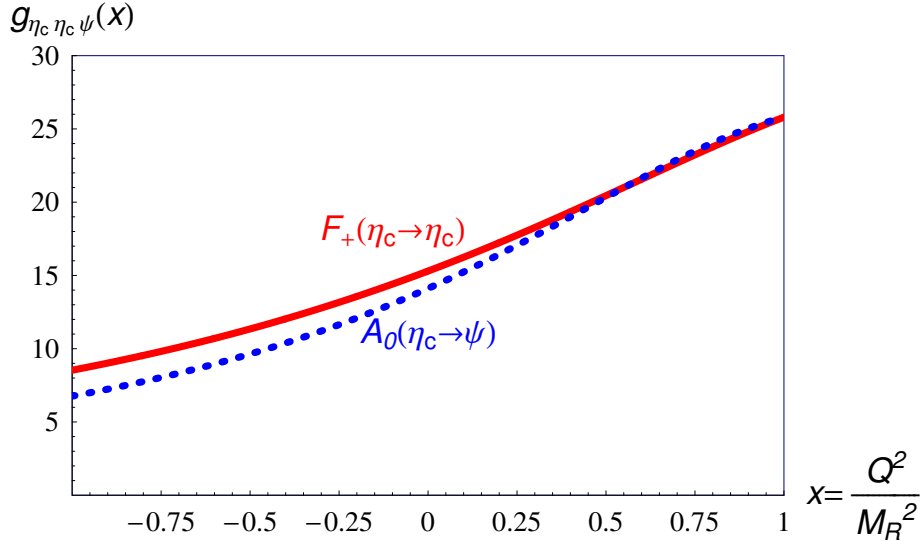}
\caption{Dependence on $x\equiv\frac{q^2}{M_R^2}$ of the
``off-shell strong coupling'' $g_{\eta_c\eta_c\psi}(x)$:
$g_{\eta_c\eta_c\widehat\psi}(x),$ from the transition
$\eta_c\to\eta_c$ (resonance $R=J/\psi,$ \textcolor{red}{red}),
versus $g_{\eta_c\widehat{\eta_c}\psi}(x),$ from the transition
$\eta_c\to J/\psi$ (resonance $R=\eta_c,$
\textcolor{blue}{blue}).}\label{Fig:SCeta}\end{center}\end{figure}

As one
example representative of the general situation, we inspect the
strong couplings of~three charmonia \cite{LMSS}. More precisely,
we start with the strong coupling $g_{\eta_c\eta_c\psi}$ of one
vector $J/\psi$ meson to two pseudoscalar $\eta_c$ mesons; the
latter strong coupling enters in and therefore can be extracted
from the residue $\mbox{Res}\,F_+^{\eta_c\to\eta_c}(M_\psi^2)$ of
the pole at $q^2=M_\psi^2$ of the form factor
$F_+^{\eta_c\to\eta_c}(q^2)$ for the (``elastic'') transition
$\eta_c\to\eta_c$ \emph{provoked\/} by the vector quark current
$\bar c\,\gamma_\mu\,c$ which couples with decay constant $f_\psi$
to the vector-meson resonance $J/\psi$ and from the residue
$\mbox{Res}\,A_0^{\eta_c\to\psi}(M_{\eta_c}^2)$ of the pole at
$q^2=M_{\eta_c}^2$ of the form factor $A_0^{\eta_c\to\psi}(q^2)$
for the transition $\eta_c\to J/\psi$ \emph{effectuated\/} by the
axial-vector quark~current $\bar c\,\gamma_\mu\,\gamma_5\,c$ which
couples with decay constant $f_{\eta_c}$ to the
pseudoscalar-meson~resonance~$\eta_c.$ These read\begin{align*}
\mbox{Res}\,F_+^{\eta_c\to\eta_c}(M_\psi^2)&=g_{\eta_c\eta_c\psi}
\,\frac{f_\psi}{2\,M_\psi}\ ,\\\mbox{Res}\,A_0^{\eta_c\to\psi}
(M_{\eta_c}^2)&=g_{\eta_c\eta_c\psi}\,\frac{f_{\eta_c}}{2\,M_\psi}\
.\end{align*}Upon determination of all needed meson wave-function
parameters $\beta_{P,V}$ by requiring the dispersion
representation of the decay constants $f_{P,V}$ to reproduce their
(observed) values, the strong couplings may be calculated,
individually for each transition of interest, from the spectral
representation of the relevant form factors: our couplings'
off-resonance behaviour exhibits excellent
agreement~(Fig.~\ref{Fig:SCeta}).

\noindent Recalling $F_+^{\eta_c\to\eta_c}(0)=1,$ a combined fit
with the four parameters $g_{\eta_c\eta_c\psi},$
$A_0^{\eta_c\to\psi}(0),$ $\sigma_1^{F,\,A}$ yields~\cite{LMSS}
$$g_{\eta_c\eta_c\psi}=25.8\pm1.7\ .$$Proceeding along a similar
path, we predict \cite{LMSS} for the strong coupling
$g_{\eta_c\psi\psi}$ of one pseudoscalar~$\eta_c$ meson to two
vector $J/\psi$ mesons (which turns out to enter in only a single
meson--meson transition)$$g_{\eta_c\psi\psi}=(10.6\pm1.5)\;{\rm
GeV}^{-1}\ .$$

\section{Strong couplings of the charmonia $\eta_c$ or $J/\psi$ to
the charmed mesons $D_{(s)}$ and $D_{(s)}^*$}
\begin{figure}[h]\begin{center}
\includegraphics[scale=.56395]{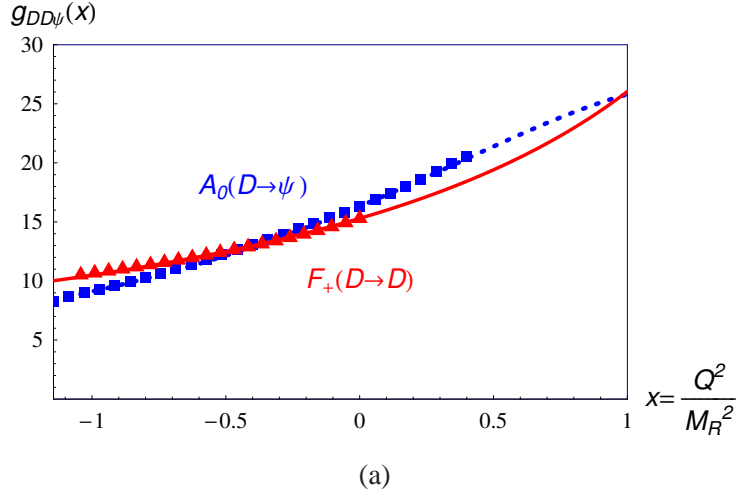}\\(a)\\
\includegraphics[scale=.56395]{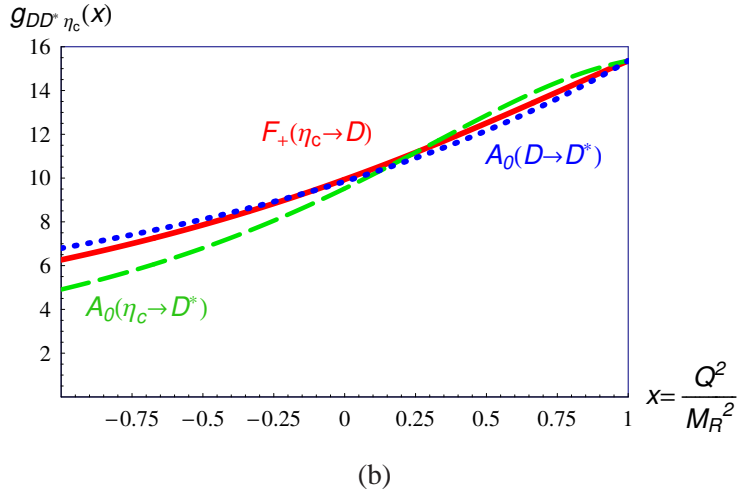}\\(b)
\caption{Dependence on $x\equiv\frac{q^2}{M_R^2}$ of ``off-shell
strong couplings'': (a)
$g_{DD\widehat\psi}(x)=\frac{2\,M_\psi}{f_\psi}\,(1-x)\,F_+^{D\to
D}(q^2)$~(\textcolor{red}{red}) or $g_{D\widehat
D\psi}(x)=\frac{2\,M_\psi}{f_D}\,(1-x)\,A^{D\to\psi}_0(q^2)$
(\textcolor{blue}{blue}), and, similarly, (b)
$g_{D\widehat{D^*}\eta_c}(x)$ (\textcolor{red}{red}),
$g_{DD^*\widehat{\eta_c}}(x)$ (\textcolor{blue}{blue}) or
$g_{\widehat DD^*\eta_c}(x)$ (\textcolor{green}{green}). For
clarity, a circumflex over the particle's symbol serves to
identify the respective resonance meson.}\label{Fig:SCD}
\end{center}\end{figure}

Allowing also for
participation of quark currents with a $d$ or $s$ quark in the
game and ``merging'' strong-coupling multiple occurrences (all of
them showing nearly perfect concord, Fig.~\ref{Fig:SCD})
yields~\cite{LMSS}

\begin{itemize}\item for the strong couplings of
the charmonia $J/\psi$ or $\eta_c$ to the non-strange charmed
mesons~$D^{(*)},$\begin{align*}&g_{DD\psi}=26.04\pm1.43\ ,
&&g_{DD^*\psi}=(10.7\pm0.4)\;{\rm GeV}^{-1}\ ,\\
&g_{DD^*\eta_c}=15.51\pm0.45\ ,
&&g_{D^*D^*\eta_c}=(9.76\pm0.32)\;{\rm GeV}^{-1}\
,\end{align*}\item and, for the strong couplings of the charmonia
$J/\psi$ or $\eta_c$ to the strange charmed mesons~$D_s^{(*)},$
\begin{align*}&g_{D_sD_s\psi}=23.83\pm0.78\ ,
&&g_{D_sD_s^*\psi}=(9.6\pm0.8)\;{\rm GeV}^{-1}\ ,\\
&g_{D_sD_s^*\eta_c}=14.15\pm0.52\ ,
&&g_{D_s^*D_s^*\eta_c}=(8.27\pm0.37)\;{\rm GeV}^{-1}\
.\end{align*}\end{itemize}

\section{Observations, comparison, and conclusions}Our present
study of the interaction strengths parametrizing the strong
couplings of all possible three-particle combinations of charmonia
and charmed mesons produced a variety of unexpected, if not even
astounding observations: First, the successful extrapolation of
our interpolated findings for strong couplings obtained at $q^2<0$
confirms the existence of the poles expected for $q^2>0.$ Second,
concerning SU(3) breaking, the net result of replacing the $d$
quark by an $s$ quark is a reduction of the affected strong
coupling by some 10\%. Third, despite undeniable similarities of
the approaches, our $D_{(s)}^{(*)}$ couplings \cite{LMSS} are more
than twice as large as those emerging from~QCD~sum~rules
(cf.~Table~\ref{Tab:SCP}).

\begin{table}[h]\begin{center}\caption{Confrontation of our
strong-coupling predictions with corresponding QCD sum-rule
figures \cite{Mat05,Bra14,Bra15}.}\label{Tab:SCP}\vspace{2ex}
\begin{tabular}{lcccc}\hline\hline\\[-2.7ex]Strong coupling&
$g_{DD\psi}$&$g_{DD^*\psi}$ (GeV$^{-1}$)&$g_{D_sD_s\psi}$&
$g_{D_sD_s^*\psi}$ (GeV$^{-1}$)\\[.5ex]\hline\\[-2.7ex]This
investigation \cite{LMSS}\ &$26.04\pm1.43$&$10.7\pm0.4$
&$23.83\pm0.78$&$9.6\pm0.8$\\QCD sum rules\ &$11.6\pm1.8$
\cite{Mat05}&$4.0\pm0.6$ \cite{Mat05}& $11.96\pm1.34$
\cite{Bra14}&$4.30\pm1.53$ \cite{Bra15}\\[.4ex]
\hline\hline\end{tabular}\end{center}\end{table}

\end{document}